\documentclass [12pt,preprint]{aastex}
\usepackage{epsfig}
\usepackage{natbib}

\begin{document}

\voffset-0.5cm
\newcommand{\gsim}{\hbox{\rlap{$^>$}$_\sim$}}
\newcommand{\lsim}{\hbox{\rlap{$^<$}$_\sim$}}

\title{On The Evolution Of The Spectral Break\\ 
       In The Afterglow Of Gamma Ray Bursts}

\author{Shlomo Dado\altaffilmark{1} and Arnon Dar\altaffilmark{1}}

\altaffiltext{1}{Physics Department, Technion, Haifa 32000, Israel}

\begin{abstract} 

The temporal evolution of the spectral break in the time resolved spectral 
energy density of the broad band afterglow of gamma ray bursts (GRBs) 
091127 and 080319B was shown recently to be inconsistent with that 
expected for the cooling break in the standard fireball model of GRBs. 
Here we show that it is, however, in good agreement with the predicted 
temporal evolution of the smooth injection break/bend in the cannonball 
model of GRBs.
\end{abstract}


\maketitle

\section{introduction}

Before the launch of the Swift satellite, the standard fireball (FB) model 
of gamma ray bursts (GRBs) have been widely accepted as the correct model 
of GRBs and their afterglows (see, e.g., the reviews by Meszaros~2002 
Zhang \& Meszaros~2004, Piran~2004). The rich data on GRBs obtained in 
recent years with the Swift satellite, complemented by data from 
ground-based rapid response telescopes and large follow-up telescopes, 
however, have challenged this prevailing view\footnote{For instance: (a) 
The prompt optical emission observed in several bright GRBs, such as 
990123 (Akerlof et al. 1999) and 080319B (Racusin et al.~2008), is orders 
of magnitude larger than that expected from extrapolating the prompt 
$\gamma$/X-ray emission to the optical band, it lags 
significantly after the $\gamma$/X-ray emission and its spectral and 
temporal behaviours appear uncorrelated to the prompt $\gamma$/X-ray 
emission, contrary to what is expected if the prompt emission were 
synchrotron radiation produced in FB driven shocks, internal or external 
(Kumar and Narayan 2009). (b) The fast decay and rapid 
spectral softening ending the prompt gamma-ray and X-ray emission in 
canonical GRBs cannot be explained simultaneously as high latitude 
emission from a relativistic fireball (Zhang et al. 2007). (c) The 
afterglows (AGs) of the Swift GRBs were found to be chromatic at early 
time (Covino et al.~2006) and to have chromatic breaks (Panaitescu et 
al.~2006) and not achromatic as expected in the standard FB model 
(Sari et al. 1999). Moreover, the breaks were found to be extremely 
rare  (Burrows and 
Racusin 2007) and very few satisfied all the criteria of the jet break of 
the standard FB model (Liang et al. 2008). (d)
The late X-ray afterglow in $\sim$50\% of 107 GRBs measured by Swift XRT 
and analyzed by  Willingale et al.~(2007) were found to
have decay parameters and $\gamma$/X-ray spectral indices that do not 
conform 
to the FB model closure relations (Sari et al.~1998).  
(e) The steep rise and strong spectral evolution 
of the late-time rebrightening observed in the bright optical and 
NIR afterglows of several GRBs, such as 080129 (Nardini et 
al.~2011) and 080219 (Greiner et al.~2009),  
could not be explained in the framework of the standard 
forward shock afterglow model. 
(f) Most of the well established 
correlations among GRB observables, such as the Amati relation and the 
newly discovered correlation by Liang et al. (2010), could not be 
explained by the FB model.}.  
These data from continuous observations across the 
electromagnetic spectrum of the prompt GRB and its afterglow emission in 
several bright GRBs, allowed critical tests of the 
standard FB model, despite the multiple choices, adjustable parameters and 
free parametrizations that are commonly employed in FB modelling of GRBs 
and their afterglows. 
One such test, a comparison between the observed and 
predicted temporal evolution of the spectral break in the spectral energy 
density (SED) of the afterglow of GRBs has been recently reported for the 
naked eye GRB 080319B (see, e.g., the supplementary materials in Racusin 
et al. 2008) and for GRB 091127 (Filgas et al. 2011). In both cases, the 
authors found that the spectral break is difficult to reconcile with the 
standard FB model of GRBs.

In standard fireball models, GRB afterglows arise from synchrotron 
emission of electrons accelerated in a strong shock driven into the 
circumburst medium by a highly relativistic fireball 
(Meszaros \& Rees 1997) or a conical jet. The energy 
spectrum of the synchrotron emission 
from high energy electrons accelerated by the Fermi mechanism or by 
collisionless shocks is usually modeled by several power-laws smoothly 
connected at characteristic break frequencies (see, e.g., Meisenheimer et 
al. 1989, Longair 1992, and references therein). Such breaks include a 
self absorption break at a frequency $\nu_a$ (the frequency above which 
the medium becomes optically thin to its synchrotron radiation), the 
characteristic synchrotron frequency $\nu_m$ (the peak frequency for the 
minimal energy of the radiating electrons), the break frequency $\nu_b$ 
(the peak frequency emitted by the electrons whose radiative cooling rate 
equals their acceleration/injection rate) and the cutoff frequency $\nu_c$ 
(the peak frequency emitted by the electrons with the maximal energy, 
i.e., by the electrons whose acceleration rate becomes slower than their 
energy-loss rate). Such a spectrum, which has been adopted in the standard 
FB model, predicts time-dependent breaks in the spectral energy density of 
the GRB afterglow when the break frequencies pass through the observed 
bands (Sari et al. 1998). In particular, for a spectral index $p_e<2$ of 
the radiating electrons and a density profile $n\propto r^{-k}$ of the 
circumburst medium, the standard FB model predicts an observed temporal 
dependence of the cooling break frequency $\nu_b\propto t^x$ where 
$x=(3k-4)/[2(4-k)]$. The passage of such breaks through the observed bands 
have been, however, difficult to identify reliably in the afterglows of 
most GRBs.

A successful measurement of the temporal evolution of the spectral break 
in the time-resolved SED of the afterglow of a GRB has been reported for 
the naked-eye 
burst 080319B in the supplementary information in Racusin et al.~2008. Due 
to the enormous brightness of this burst, the authors were able to fit 
broad-band SEDs at several epochs using Swift UVOT and XRT data, as well 
as a multitude of optical and NIR ground-based data. They measured a 
temporal evolution of the spectral break with a temporal index 
$x=1.08\pm0.04$ for $t<1800$s and $x=-1.00 \pm 0.14$ for the late 
($t>1800$ s) afterglow, while a wind profile with $k=2$ yields $x=0.5$ and 
a constant ISM density, i.e., $k=0$, yields $x=-0.5$. Racusin et al. have 
also examined both pre- and post-jet break FB closure relations with and 
without lateral spreading, but were unable to reproduce the observed 
temporal behaviour with the standard fireball model. They concluded that 
the temporal behaviour of the spectral break is difficult to reconcile 
with the standard FB model of GRBs where a single jet that propagates in a 
complex density medium produces the GRB afterglow through synchrotron 
radiation. They suggested that perhaps with further modifications to 
micro-physical parameters and model dependencies, one could manufacture a 
standard FB model that could adequately reproduce the observed data. 
Alternatively, they found that the chromatic behaviour of the broadband 
afterglow of GRB 080319B could be explained by viewing the GRB down the 
very narrow inner core of an assumed two-component jet that is expanding 
into a wind-like environment.

A second measurement of the temporal evolution of the spectral break in 
the time resolved SED of the late afterglow of a GRB was reported recently 
for GRB 091127 in Filgas et al. (2011). The accurate and well sampled 
measurements of the X-ray and optical/NIR multi-color light curves of its 
afterglow showed evidence of a spectral break moving from high to lower 
energies with increasing time.  Detailed fitting of the time-resolved SED 
showed that the break is very smooth and evolved with a temporal index 
$x=-1.23\pm 0.06$, inconsistent with  either $x=0.5$ (wind profile)  
or $x=-0.5$ (ISM profile) that was expected from the 
standard FB model. Filgas et al. concluded that their analysis provides 
further evidence that the standard fireball model is too simplistic, and 
time-dependent micro-physical parameters may be required to model the 
growing number of well-sampled afterglow light curves, which is beginning 
to place strong constraints on the fireball model of GRBs and on possible 
alternatives.

In this letter, however, we show that the observed temporal evolution of 
the spectral break in the time-resolved SED of the afterglow of GRBs 
080319B and 091127 is in good agreement with that predicted for the smooth 
injection break/bend of the cannonball (CB) model of GRBs (Dar \& De 
R\'ujula 2004, Dado et al. 2009).

\section{Temporal behaviour of the  break frequency in the CB model}

In the CB model (e.g., Dar \& De R\'ujula~2004, Dado et al.~2009
Dado \& Dar 2009a, and 
references therein), {\it long-duration} GRBs and their afterglows are 
produced by bipolar jets of highly relativistic CBs (Shaviv \& Dar 1995, 
Dar 1998, Dar \& Plaga 1999, Dar \& De R\'ujula 2000, Dado et al. 2002) 
ejected in supernova explosions (Dar et al. 1992, Dar 1999, Dar \& Plaga 
1999, Dar \& De R\'ujula 2000, Dado et al. 2002).  The prompt MeV $\gamma$ 
and keV X-ray emission is dominated by inverse Compton scattering (ICS) of 
glory light (progenitor light scattered by the ejecta/winds blown from the 
progenitor star before the GRB). The ICS is overtaken later by the 
synchrotron radiation (SR), which begins slightly after the beginning of 
the prompt GRB, when the CB enters the pre-supernova wind/ejecta of 
the progenitor star (see, e.g., Dado et al. 2009).  
The early time optical/NIR emission is dominated by this SR and may have a 
significant contribution also from bremsstrahlung. Thus, the prompt and 
early time keV-MeV emission does not extrapolate smoothly to the optical 
emission, and the broad band light curves show chromatic behaviour and 
complex spectral evolution, as was observed in several GRBs such as
990123, including GRB 
080319B (e.g., Racusin et al. 2008, Bloom et al. 2009). In such cases, a 
spectral break that is inferred from extrapolating the early time X-ray 
SED to a lower energy and the SED of the early time NIR/optical emission 
to a higher energy is neither an SR cooling break nor an injection 
break/bend. Only when the SR emission dominates the entire broad-band 
afterglow can one test whether the broad band SED has a break consistent 
with the cooling break of the FB model or the smooth injection break/bend 
predicted by the CB model.

The spectral and temporal behaviour of the afterglow  
(the radiation emitted after the fast decline of the prompt keV-MeV 
emission) that is predicted by the CB model is much simpler. 
In the CB model, 
the circumburst medium in front of a CB is completely ionized 
by the CB's radiation. In the CB's rest frame, 
the ions of the medium that are continuously impinging on the CB 
generate within it a turbulent magnetic field, which is assumed to 
be in approximate energy equipartition, 
$B\approx \sqrt{4\,\pi\, n\, m_p\, c^2}\, \gamma$ (Dado et al. 2002),
where $n$ is the external proton density  per cm$^3$ and $m_p$ is the 
proton mass.
The electrons, which enter the CB at a time $t$ with the bulk motion 
Lorentz factor $\gamma(t)$ in its rest frame, are  
Fermi accelerated and cool rapidly by synchrotron radiation.  This SR 
is isotropic in the CB's rest frame and  has a smoothly broken power law 
with  a characteristic injection bend/break frequency $\nu'_b(t)$,  
which is the typical synchrotron frequency radiated by the ISM electrons 
that enter  the CB at time $t$ with a relative Lorentz factor $\gamma(t)$.
In the observer frame, the emitted photons are beamed 
into a narrow cone along the CB's direction of motion 
by its highly relativistic bulk motion, their arrival times 
are aberrated and their energies are boosted by
its  bulk motion Doppler factor $\delta$ and redshifted by the cosmic
expansion during their  travel time to the observer.

In particular, 
in the observer frame (see, e.g. Eq.~(25) in Dado et al.~2009),
\begin{equation}
\nu_b(t)=\delta(t)\,\nu'_b(t)/(1+z)\propto n^{1/2}\,[\gamma(t)]^3 
\delta(t)\,,
\label{nub}     
\end{equation}
where $n$ is the circumburst density, and the spectral 
energy density of the {\it unabsorbed} X-ray afterglow 
has the form (see, 
e.g., Eq.~(26) in Dado et al.~2009),
\begin{equation}
F_{\nu} \propto  n^{(\beta+1)/2}\,
[\gamma(t)]^{3\,\beta-1}\, [\delta(t)]^{\beta+3}\, \nu^{-\beta}\, ,
\label{Fnux}
\end{equation}
where $\beta+1=\Gamma$ is the photon spectral index of the 
emitted (unabsorbed) radiation.

At an early time before the CB has swept in a relativistic mass 
comparable to its mass, $\gamma$ and $\delta$ change rather slowly 
as a function of time and stay put at their initial values.  Hence, 
at an early time, the distance from the progenitor star to the CB 
is given approximately by 
$r\approx\gamma(0)\, \delta(0) t/(1+z)\propto t$
where $t$ is the photon arrival time in the observer frame
after the beginning of the burst. 
Consequently, at an early time,
\begin{equation}
\nu_b\propto n(r[t])^{1/2}\,.
\label{nubent}
\end{equation}
Hence, for $r\propto t$, a circumburst density profile $n\propto r^{-k}$ 
yields
\begin{equation}
\nu_b\propto t^{-k/2}\,,
\label{nube}
\end{equation}
e.g., $\nu_b(t)$ rises like $t$ for $k=-2$, behaves like a
constant for a constant density ($k=0$), and declines like $t^{-1}$ for a
wind profile ($k=2$).

In the CB model, Eq.~(3) is valid also when a CB crosses a 
density bump, provided that the relativistic mass
swept in by the CB 
during this crossing is small compared to its mass. Then,
quite generally, for both a wind profile and a density bump, 
roughly, $\nu_b(t)\propto [\nu^{\beta}\,F_\nu(t)]^{1/(1+\beta)}$.

For $\gamma\gg 1$ and  small viewing angles $\theta\ll 1$, 
to a good approximation 
$\delta(t)=1/\gamma(t)\,(1-cos\theta)\approx 2\, 
\gamma(t)/(1+\gamma(t)^2\, \theta^2)$.
The decelaration of the CBs during the afterglow phase 
decreases  $\gamma(t)$ and  
$\gamma(t)^2 \theta^2$ becomes $\ll 1$ yielding
$\delta(t)\simeq 2\,\gamma(t)$,
which simplifies
Eqs.~(2) and (3) to 
\begin{equation}
F_{\nu} \propto  n^{\Gamma/2}\,[\gamma(t)]^{4\,\Gamma-2}\,,
\label{Fnuxl}
\end{equation}
\begin{equation}
\nu_b \propto  n^{1/2}\,[\gamma(t)]^4 \,,
\label{nubl}
\end{equation} 
respectively. Consequently, for an unabsorbed X-ray afterglow of the form 
$F_{\nu}(t)\propto t^{-\alpha}\nu^{-\beta}$ with $\beta=\Gamma-1$,
the CB model predicts for a constant density ISM ($k$=0)
that as soon as $\delta \simeq 2\,\gamma$,
\begin{equation} 
\nu_b(t)\propto t^{-\alpha/(\Gamma-1/2)}\,. 
\label{nubl1}
\end{equation} 
In ordinary GRBs where $\gamma(0)\,\theta \sim 1$,
this happens quite early  and somewhat later 
in X-ray flashes that are far off-axis GRBs.

\section{Comparison with experiment}

{\bf GRB 091127:} In Fig.~1a we compare the 0.3-10 keV X-ray lightcurve of 
the afterglow of GRB 091127 that was measured with the Swift X-ray 
telescope (XRT) and reported in the Swift XRT light curve repository 
(Evans et al.~2007,2009) and its CB model description (Dado \& Dar 2009b) 
assuming a constant circumburst density. The agreement is quite 
satifactory ($\chi^2/DOF=360/342=1.05$). The asymptotic decline of the 
light curve is well represented by the power law $F_\nu(t)\propto 
t^{-\alpha}$ with $\alpha=1.6\pm 0.04$. The spectral analysis of the Swift 
XRT data by Filgas et al.~(2011) yielded a photon spectral 
index $\Gamma_X=1.748 \pm 0.004$. Hence, for an assumed constant ISM 
density along the GRB trajectory, the CB model predicts a power-law index 
$x=-\alpha/(\Gamma-1/2)=-1.28 \pm 0.04$ for the temporal decay beyond 
$\sim$4000 s of the injection bend/smooth break in the SED of GRB 091127, 
in good agreement with the reported best fit value $x=-1.23\pm 0.06$ 
(Filgas et al. 2011). This is demonstrated in Fig.~1b where we compare the 
measured temporal decay of the smooth break and that predicted by the CB 
model (Eq.~(1)) and its approximate power-law behaviour (Eq.~(7)). The 
normalizations were adjusted to fit the data. The CB model fits have 
$\chi^2/DOF=0.76$ and $\chi^2/DOF=0.62$, respectively.

{\bf GRB 080319B}: The observed X-ray and optical light curves of GRB 
080319B are well reproduced with the CB model (Dado \& Dar 2008). This is 
demonstrated for the X-ray light curve in Fig.~1c. The mean photon 
spectral index in the X-ray band that was inferred from the X-ray 
observations of GRB 080319B with the Swift/XRT (Evans et al. 2007,2009)  
in the PC mode was $\Gamma=1.82\pm 0.06 $ for $t>4000$ s. For such a 
photon spectral index and a wind density profile $n\propto r^{-2}$, the CB 
model predicts (e.g., Dado et al. 2009) an SED with a temporal decline 
$F_\nu(t)\propto t^{-\Gamma}=t^{-1.82 \pm 0.06}$ which is in good 
agreement with the temporal decline $F_\nu(t)\propto t^{-1.85\pm 0.05}$ 
observed with the Swift/XRT during the time interval $1800<t<40000$ s.  
For a wind profile ($k=2$), Eq.~(1) predicts a temporal decline of the 
injection break/bend, $\nu_b(t)\propto t^{-1}$, i.e., a temporal index 
$x=-1$. Around $t=40000$ s the decline of the SED that was 
observed by Swift/XRT changed to $F_\nu(t)\propto t^{-1.28 \pm 0.04}$. In 
the CB model this was interpreted as due to the CB entering around that 
time a constant ISM density, for which the CB model predicts an index 
$x=-\alpha/(\Gamma-1/2)=(1.28\pm 0.04)/(1.32\pm 
0.06)=-0.97 \pm 0.07$ (see Eq.~(7) for the temporal decline of the 
spectral break/bend). Both values are in good agreement with the mean 
value $x=-1.0 \pm 0.14 $ measured by Racusin et al. 2008 for 
$t>1800$ s as shown in Fig.~1d.

\section{Conclusion}

Convincing measurements of the temporal evolution of the smooth spectral 
break/bend in the time-resolved spectral energy density of the broad band 
late afterglow of gamma ray bursts were recently reported for GRBs 091127 
and 080319B. These measurements allow a critical test of GRB models, such 
as the standard fireball model and the cannonball model, because the 
parameters which determine the spectral evolution in these models are 
constrained by closure relations. While the prediction of the standard 
fireball model cannot be reconciled with the measured temporal evolution 
of the smooth spectral break/bend in these GRBs, the predictions of the 
cannonball model are in good agreement with these measurements.

\noindent
{\bf Acknowledgment:} 
We thank an anonymous referee for useful comments.
This work made use of data supplied by the UK Swift Science Data Centre 
at the University of Leicester.

\newpage
\begin{figure}[]
\centering
\vspace{-2cm}
\vbox{
\hbox{
\epsfig{file=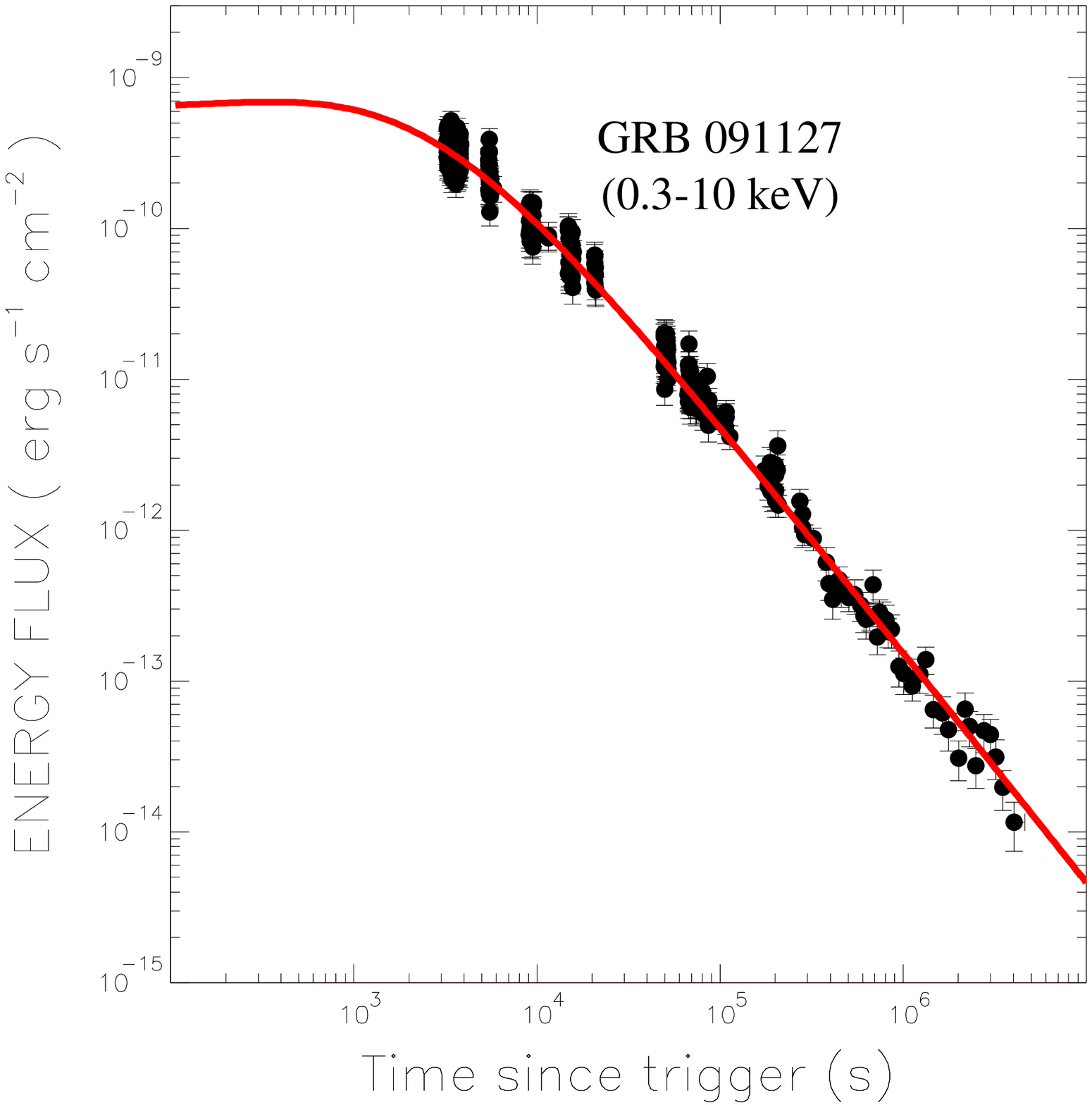,width=8.cm,height=8.cm}
\epsfig{file=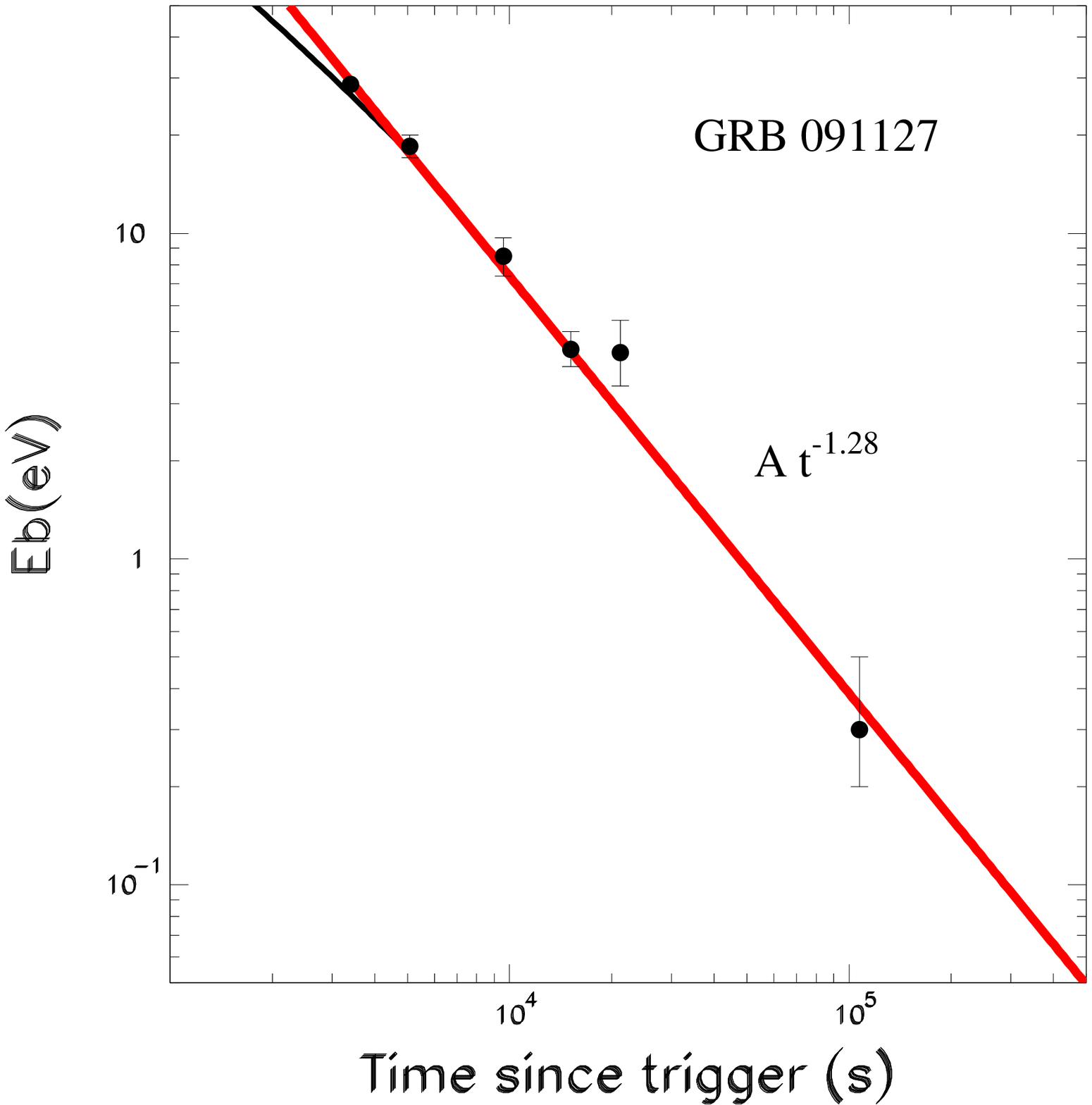,width=8.cm,height=8.cm}
}}
\vbox{
\hbox{  
\epsfig{file=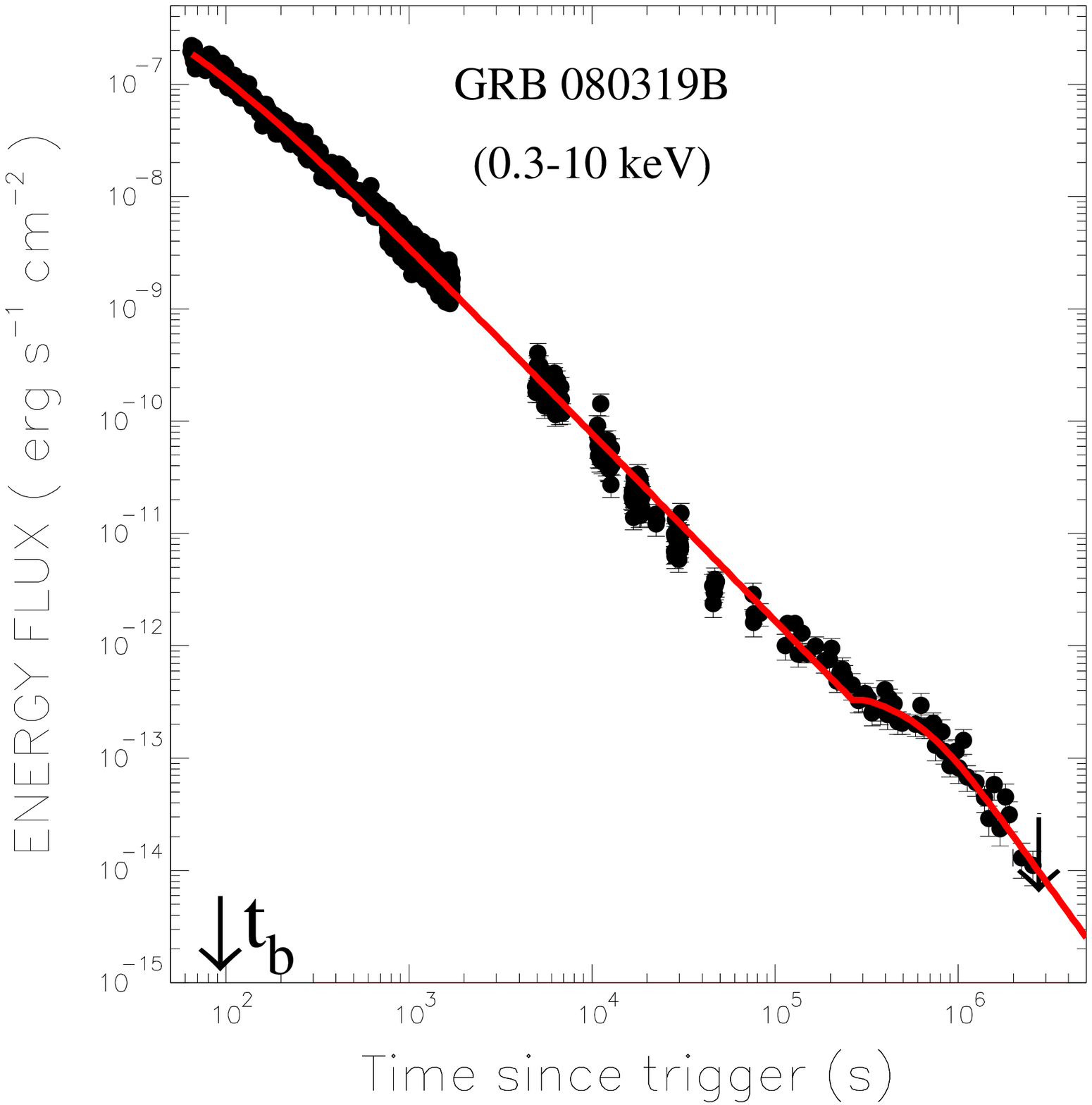,width=8.cm,height=8.cm}
\epsfig{file=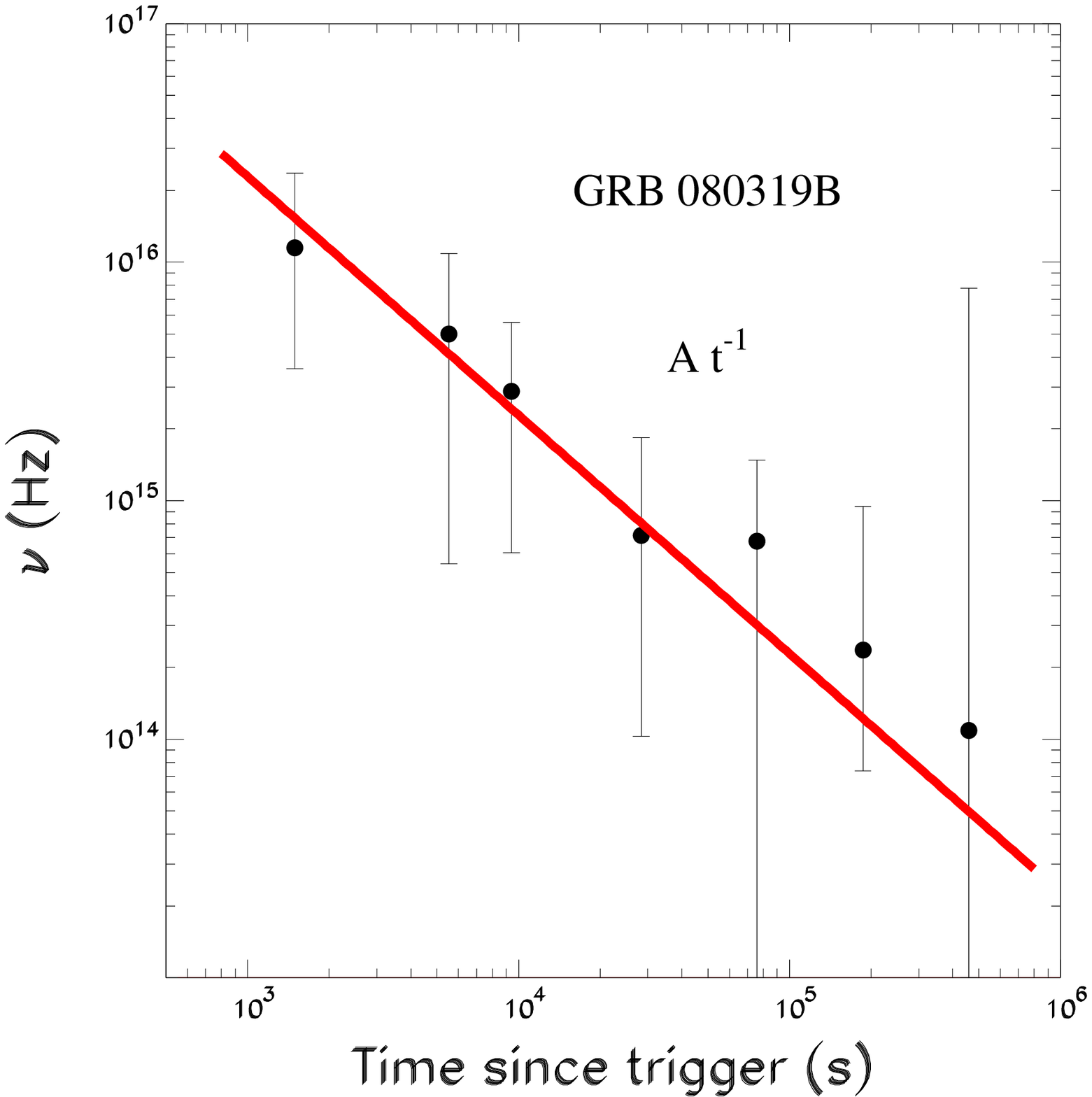,width=8.cm,height=8.cm}
}}
\caption{
{\bf Top left:} Comparison between the 0.3-10 keV light curve
of the X-ray afterglow of GRB 091127 measured with Swift and reported
in the 
Swift XRT lightcurve repository (Evans et al. 2007, 2009)) and its CB 
model
description (Eq.~(2)). For details see Dado \& Dar 2009b.
{\bf Top right:} Comparison between the temporal evolution
of the smooth spectral break/bend of the time resolved SED in GRB 091127
as measured by Filgas et al. 2011 and the power-law predicted by the CB
model (Eq.~(7).
{\bf Bottom left:} Comparison between the X-ray light curve
of GRB 080319B as measured with the Swift XRT
(Evans et al. 2007, 2009)) and its CB model
description (Eq.~(2)), (for details see Dado \& Dar 2008).
{\bf Bottom right:}
Comparison between the temporal evolution
of the spectral break of the time resolved broad band SED
of the late afterglow in GRB 080319B
as inferred by Racusin et al. 2008 and that
predicted by the CB model (Eq.~(7))}.
\label{FIG01}
\end{figure}

\end{document}